\documentclass[slac_one]{revtex4}
\usepackage{graphicx}
\usepackage{fancyhdr}
\newcommand{\Lumint}{{\cal L}_{\rm int}}
\begin{document}
\title{{\small{2005 ALCPG \& ILC Workshops - Snowmass,
U.S.A.}}\\ 
\vspace{12pt} \begin{center} Discriminating Large Extra
Dimensions at the ILC with Polarized Beams
\end{center}
} 
%
\author{A. A. Pankov, A. V. Tsytrinov}
\affiliation{Pavel Sukhoi Technical University, Gomel 246746,
Belarus}
\author{N. Paver}
\affiliation{University of Trieste and INFN, 34100 Trieste, Italy}
\begin{abstract}
Non-standard scenarios described by effective interactions can manifest
themselves indirectly, {\it via} corrections to the Standard Model cross
sections. It should be desirable to identify at a given confidence level the
source of such deviations among the different possible explanations. We
here discuss the identification reach on gravity in extra dimensions from
the four-fermion compositeness-inspired contact interactions and
{\it viceversa}, using as basic observable the differential cross section of
$e^+e^-\to\bar{f}f$ at the ILC, and emphasize
the r{\^o}le of beams polarization in enhancing the identification
sensitivity.
\end{abstract}
\maketitle
\thispagestyle{fancy}
\section{INTRODUCTION}\label{sec:intro}
New-physics scenarios (NP) based on very heavy virtual quanta exchanges can
be described, below the direct production threshold, by effective,
contact-interactions that can can have only {\it indirect} signatures by
contributing corrective terms to the Standard Model (SM) amplitudes,
suppressed by some power of the ratio between the collider c.m. energy and
the above mentioned characteristic high mass scales.
These corrections will reveal themselves {\it via}
deviations of the measured observables from the SM predictions or, in few
specific cases, by the observation of processes forbidden by the SM.
\par
In principle, different kinds of NP interactions may produce similar
deviations and, consequently, it would be desirable to assess, for each
non-standard model, not only the ``discovery reach'', represented by
the maximal value of the relevant mass scale below which a deviation can be
observed at a given C.L. within the experimental accuracy but, also,
the ``identification reach'', defined as the upper limit of the mass
range of values for which the model can be discriminated from the other
potentially competing scenarios.
\par
We will focus on the discrimination reach on the ADD models of gravity
in large, compactified, extra spatial dimensions \cite{Arkani-Hamed:1998nn},
with respect to the four-fermion contact interactions inspired by
compositeness \cite{Eichten:1998nn}, and {\it viceversa}, looking at the
differential cross sections of
\begin{equation}
e^++e^-\to \bar{f}+f,
\label{proc}
\end{equation}
with $f=l,q$ ($l=\mu,\,\tau$; $q=c,\,b$), at the ILC with longitudinally
polarized beams \cite{Abe:2001nn,Moortgat-Pick:2005cw}.
In Ref.~\cite{Pasztor:2001hc}, the identification reach on individual
contact-interactions was studied by applying a Monte Carlo technique to
lepton-pair production with unpolarized beams. An approach based on the
polarized differential distributions for lepton pair production processes
was proposed in Ref.~\cite{Pankov:2005kd}. We here discuss the benefits of
longitudinal beams polarization in improving the identification reaches and
consider also quark-pair production channels.
\section{DIFFERENTIAL CROSS SECTIONS AND DEVIATIONS FROM THE SM}
\label{sec:cross}
Neglecting all fermion masses with respect to the c.m. energy
$\sqrt s$, the polarized differential cross section of processes
(\ref{proc}) can expressed as follows \cite{Schrempp:1987zy}:
\begin{equation}
\frac{d\sigma^{\rm pol}}{d z}=
\frac{1}{4}\left[\left(1-P_1\right)\left(1+P_2\right)
\left(\frac{d\sigma_{\rm LL}}{d z} +\frac{d\sigma_{\rm LR}}{d z}\right)+
\left(1+P_1\right)\left(1-P_2\right)
\left(\frac{d\sigma_{\rm RR}}{d z}
+\frac{d\sigma_{\rm RL}}{d z}\right)\right],
\label{crossdif-pol}
\end{equation}
where $z=\cos\theta$ is the angle between the incoming and outgoing fermions
in the c.m. frame and ($\alpha,\beta={\rm L,R}$):
\begin{equation}
\frac{d\sigma_{\alpha\beta}}{d z}=N_{\rm colors}\frac{3}{8}\sigma_{\rm pt}
\vert {\cal M}_{\alpha\beta}\vert^2\, (1\pm z)^2.
\label{helicity}
\end{equation}
$P_1$ and $P_2$ the degrees of longitudinal polarization of the electron
and positron beams, respectively, and the `$\pm$' signs apply to the cases
${\rm LL}$, ${\rm RR}$ and ${\rm LR}$, ${\rm RL}$, respectively.
\par
According to sec.~\ref{sec:intro}, the reduced helicity amplitudes appearing
in Eq.~(\ref{helicity}) can be expanded into the SM part represented by
$\gamma$ and $Z$ exchanges, plus corrections depending on the considered NP
model:
\begin{equation}
{\cal M}_{\alpha\beta}={\cal M}_{\alpha\beta}^{\rm SM}
+\Delta_{\alpha\beta}({\rm NP}).
\label{amplit}
\end{equation}
The examples explicitly considered here are the following ones:
\par
{\it a)} The ADD large extra dimensions scenario \cite{Arkani-Hamed:1998nn},
where only gravity can propagate in extra dimensions, and correspondingly
a tower of graviton KK states occurs in the four-dimensional space
\cite{Han:1998sg,Giudice:1998ck}. In the parameterization of
Ref.~\cite{Hewett:1998sn}, the ($z$-dependent) deviations can be expressed as
\cite{Cullen:2000ef}:
\begin{equation}
\Delta_{\rm LL}({\rm ADD})=\Delta_{\rm RR}({\rm ADD})=f_G(1-2z),\quad
\Delta_{\rm LR}({\rm ADD})=\Delta_{\rm RL}({\rm ADD})=-f_G(1+2z),
\label{ADD}
\end{equation}
where $f_G=\lambda\,s^2/(4\pi\alpha_{\rm e.m.}\Lambda_H^4)$, $\lambda=\pm1$, $\Lambda_H$
being a phenomenological cut-off on the integration on the KK spectrum.
\par {\it b)} Gravity in ${\rm TeV}^{-1}$--scale extra dimensions,
where also the SM gauge bosons can propagate there,
parameterized by the ``compactification scale'' $M_C$
\cite{Cheung:2001mq,Rizzo:1999br}:
\begin{equation}
\Delta_{\alpha\beta}({\rm TeV})=
-\left(Q_eQ_f+g_\alpha^e\, g_\beta^f\right)\pi^2/(3\, M_C^2).
\label{tevscale}
\end{equation}
\par
{\it c)} The four-fermion contact-interaction scenario (CI)
\cite{Eichten:1998nn} where, with $\Lambda_{\alpha\beta}$ the
``compositeness'' mass scales ($\eta_{\alpha\beta}=\pm1$):
\begin{equation}
\Delta_{\alpha\beta}({\rm CI})=
{\eta_{\alpha\beta}s}/({\alpha_{\rm e.m.}}{\Lambda^2_{\alpha\beta}}).
\label{CI}
\end{equation}
\par
In cases {\it b)} and {\it c)} the deviations are $z$-independent, whereas
in the case {\it a)} they introduce extra $z$-dependence in the angular
distributions. The consequence is that the ADD contribution to the
integrated cross sections is tiny, because the interference with the SM
amplitudes vanishes in these observables. Current experimental lower
bounds on the mass scales $M_H$ and $M_C$ are reviewed, e.g., in
Ref.~\cite{Cheung:2004ab} ($M_H>1.1-1.3\, {\rm TeV}$, $M_C>6.8\, {\rm TeV}$),
while those on $\Lambda$s, of the order of 10 TeV, are detailed in
Ref.~\cite{Eidelman}.
\section{DERIVATION OF THE IDENTIFICATION REACHES}
\label{sec:identification}
Let us assume one of the models, for example the ADD model (\ref{ADD}), to be
the ``true'' one, i.e., to be consistent with data
for some value of $\Lambda_H$. To estimate the level at which it may
be discriminated from other, in principle competing NP scenarios (``tested''
models), for any values of the relevant mass parameters, say
example one of the four-fermion CI models (\ref{CI}), we
introduce relative deviations of the differential cross section (denoted by
$\cal O$) from the ADD predictions due to the CI in each angular bin, and a
corresponding $\chi^2$ function:
\begin{equation}
{\Delta} ({\cal O})= \frac{{\cal O}({\rm CI})-{\cal O}({\rm
ADD})}{{\cal O}({\rm ADD})}; \qquad\quad
\chi^2({\cal O})= \sum_{\rm bins}\left(\frac{\Delta({\cal O})^{\rm
bin}} {\delta{\cal O}^{\rm bin}}\right)^2.
\label{chi}
\end{equation}
Here, $\delta{\cal O}$s represent the expected relative uncertainties,
which combine statistical and systematic ones, the former one being related
to the ADD model prediction. Consequently, the $\chi^2$ of Eq.~(\ref{chi})
is a function of $\lambda/\Lambda_H^4$ and the considered
$\eta/\Lambda^2$, and we can determine the
``confusion'' region in this parameter plane where also the corresponding CI
model may be considered as consistent with the ADD predictions at the chosen
confidence level, so that an unambiguous identification of ADD cannot be
made. We choose $\chi^2<3.84$ for 95\% C.L..
\par
For the numerical analysis, we consider an ILC with $\sqrt s=0.5\, {\rm TeV}$
time-integrated luminosity $\Lumint$ from $100\, {\rm fb}^{-1}$ up
to $1000\, {\rm fb}^{-1}$; reconstruction efficiencies 95\% for $l^+l^-$,
60\% for $b\bar{b}$ and 35\% for $c\bar{c}$. We divide the angular range,
$\vert z\vert<0.98$ in ten bins. To account for the major systematic
uncertainties, we assume $\delta\Lumint/\Lumint=0.5\%$, and
$\vert P_1\vert=0.8$ and $\vert P_{2}\vert=0.6$ with
$\delta P_1/P_1=\delta P_2/P_2=0.2\ \%$. Specifically, we consider the
four polarized cross sections with the configurations $(P_1,P_2)=
(0.8,-0.6)$ and $(-0.8,0.6)$,
and combine them into the $\chi^2$ also accounting for their mutual
statistical correlations.
\par
\begin{figure}[htbp] 
\centerline{
\includegraphics[width=6.3cm,angle=0]{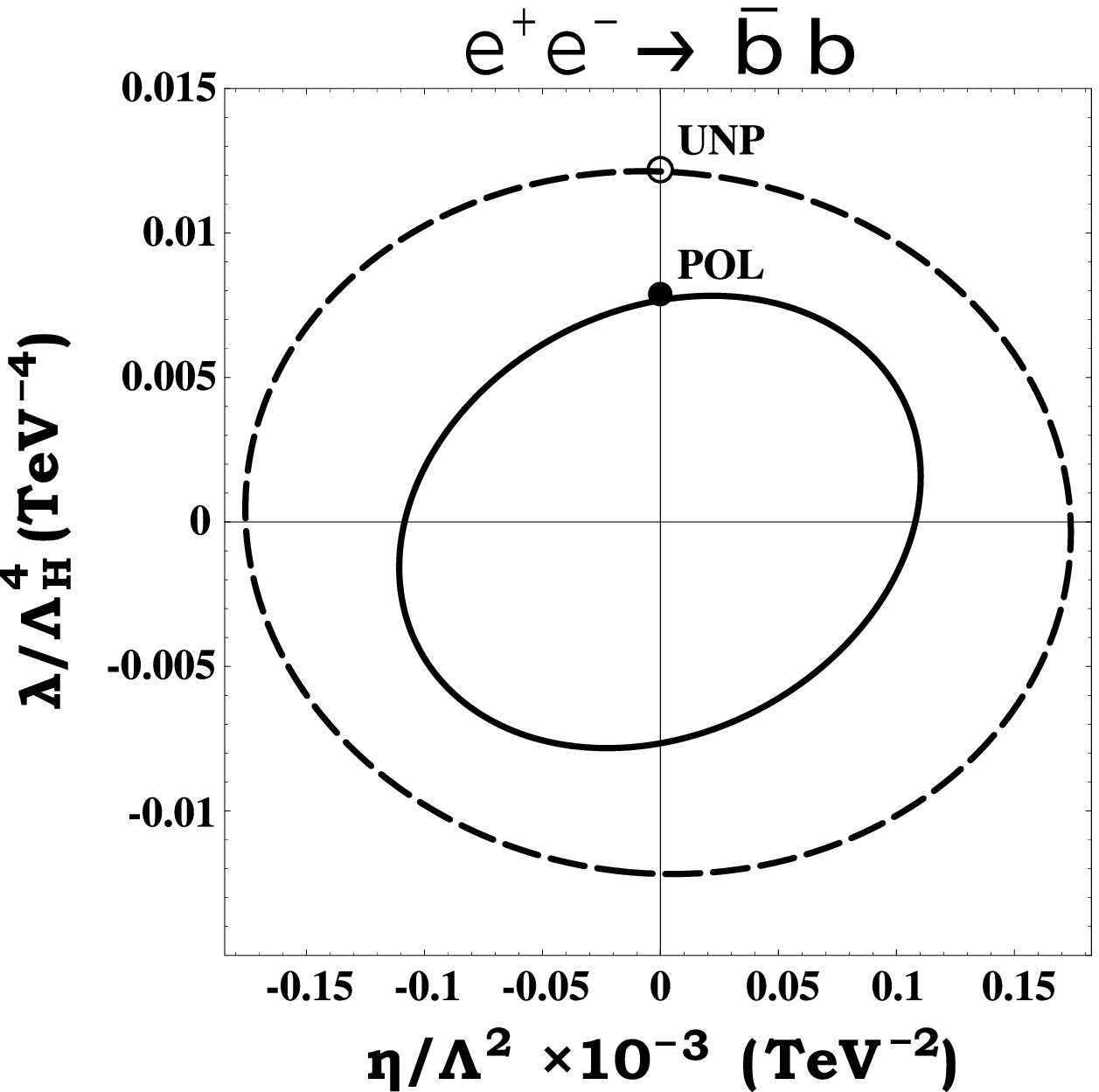}
\hspace*{2.0cm}
\includegraphics[width=8.0cm,angle=0]{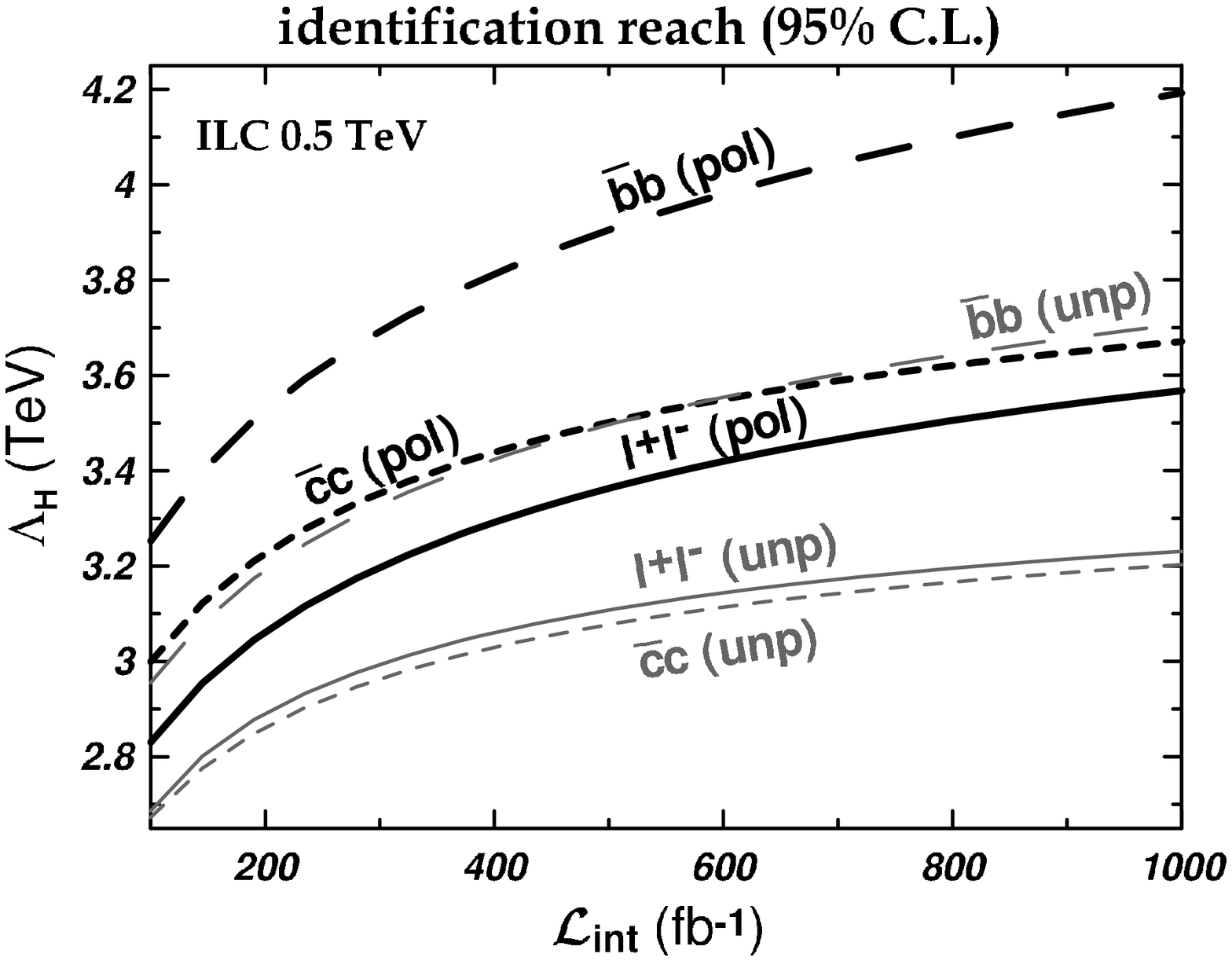}}
\caption{Left panel: confusion region (95\% C.L.) for ADD
and VV models from $e^+e^-\to\bar{b}b$ with $\Lumint=100$ fb$^{-1}$.
Right panel: identification reach on $\Lambda_H$ in the ADD model {\it vs.}
$\Lumint$ obtained from all processes (\ref{proc})
with unpolarized and polarized beams.}
\label{fig1}
\end{figure}

Fig.~\ref{fig1} (left panel) shows as an example the ``confusion region''
between the ADD and the VV models, resulting from the process
$e^+e^-\to\bar{b}b$, with the above inputs and $\Lumint=$100 fb$^{-1}$,
both for unpolarized and polarized beams.
The figure shows that a maximal absolute value of the $\lambda/\Lambda^{4}_{H}$
(equivalently, a minimal value of $\Lambda_H$) can be found, for which the
``tested'' VV model hypothesis is expected to be excluded at the 95\% C.L.
for any value of the CI parameter $\eta/\Lambda^{2}$. We denote the
corresponding ADD mass scale parameter as $\Lambda_H^{\rm VV}$
and call it ``exclusion reach'' of the VV model. The same procedure
can be applied to all other types of effective contact interaction models
considered in Eqs.~(\ref{CI}) and (\ref{tevscale}), and leads to the
corresponding ``exclusion reaches'' $\Lambda_H^{\rm AA}$, $\Lambda_H^{\rm RR}$,
$\Lambda_H^{\rm LL}$, $\Lambda_H^{\rm LR}$, $\Lambda_H^{\rm RL}$
and $\Lambda_H^{\rm TeV}$. As the final step, the
``identification reach'' on the ADD scenario can be defined as the minimum of
the $\Lambda_H$ ``exclusion reaches'',
$\Lambda_H^{\rm ID}=min\{\Lambda_H^{VV},\, \Lambda_H^{AA},
\Lambda_H^{RR},\, \Lambda_H^{LL},\, \Lambda_H^{LR},\,
\Lambda_H^{RL},\, \Lambda_H^{TeV}\}$. Clearly,
$\Lambda_H<\Lambda_H^{\rm ID}$ allows to exclude {\it all}
composite-like CI models as well as the ${\rm TeV}^{-1}$ gravity
model. The results of this kind of analysis for all processes
(\ref{proc}) with unpolarized beams  as well as polarized beams, and the
corresponding ``identification reach'' on $\Lambda_H$,
are shown in Fig.~\ref{fig1} (right panel).
\begin{figure}[htbp] 
\centerline{
\includegraphics[width=8.0cm,angle=0]{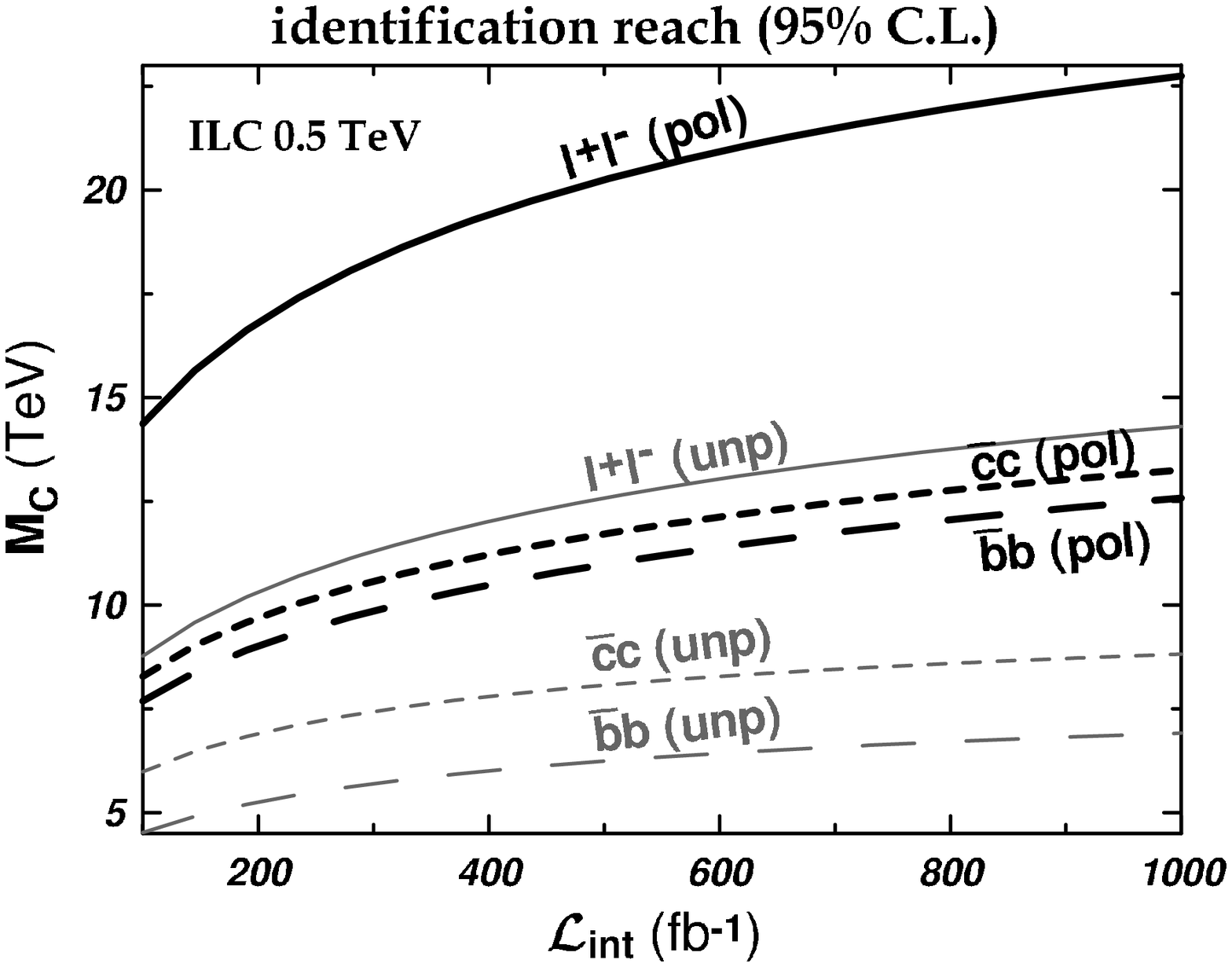}
\hspace*{0.5cm}
\includegraphics[width=8.0cm,angle=0]{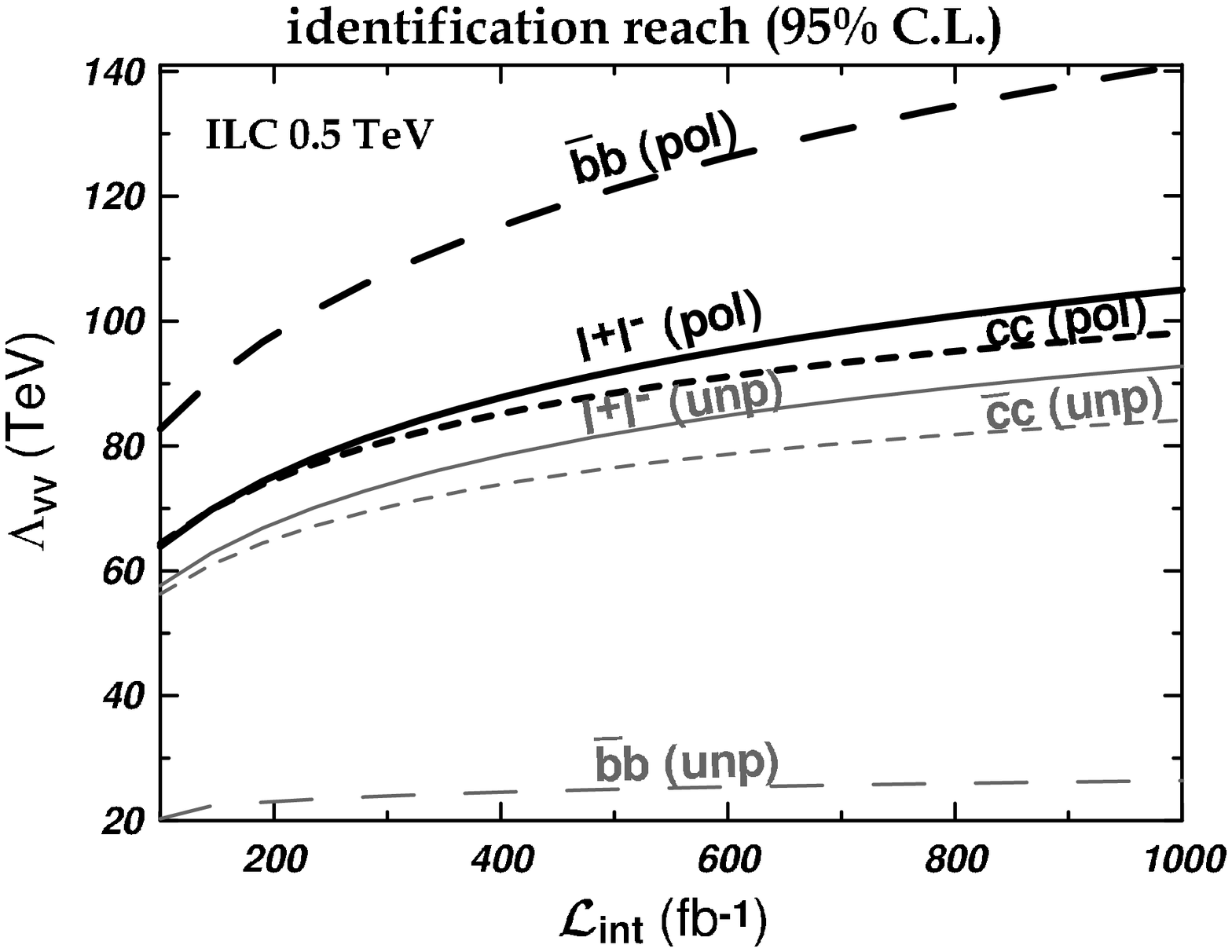}}
\caption{ 95\% {\rm CL} identification reach on the
cutoff scale $\rm{M_C}$ in the TeV model (left panel) and $\Lambda_{\rm{VV}}$
in the VV model (right panel) as a function of the
integrated luminosity obtained from the fermion pair production
processes with unpolarized and both polarized beams at ILC(0.5
TeV).}
\label{fig2}
\end{figure}
\par
The simple, $\chi^2$-based procedure outlined above can be applied in turn
to all individual processes, sorces of the corrections in Eqs.~(\ref{CI})
and (\ref{tevscale}), and distinction reaches on the relevant mass
parameters can be derived analogously. In Fig.~\ref{fig2} we show, as
examples, the results for the compactification scale $M_C$ and the
CI compositeness scale $\Lambda_{\rm VV}$. One can notice, from both
Figs.~\ref{fig1} and \ref{fig2}, the essential r{\^o}le of beam
polarization in increasing the discrimination sensitivity on the different
NP scenarios.
\par
In conclusion, we have developed a specific approach based on the
differential polarized cross sections to search for and identify
spin-2 graviton exchange with uniquely distinct signature. Fig.~\ref{fig1} (right panel)
shows that, of the three considered processes, $\bar{b}b$ pair
production process definitely has the best identification
sensitivity on the scale $\Lambda_H$ characterizing the ADD model
for gravity in ``large'' compactified extra dimensions. As one
can see, in the polarized case, the identification reach ranges
from 3.3 TeV to 4.2 TeV, depending on the luminosity.
\par

\end{document}